\documentclass[aps,prl,twocolumn,showpacs]{revtex4}

\usepackage[T1]{fontenc}
\usepackage[latin1]{inputenc}
\usepackage[english]{babel}

\usepackage{textcomp}
\usepackage{amsmath}
\usepackage{amsfonts}
\usepackage{amssymb}
\usepackage{units}

\usepackage{graphicx}
\usepackage{xcolor}

\begin{document}

\title{Anomalous Magnetic Hyperfine Structure of the $^{229}$Th
Ground-State Doublet in Muonic Atom}

\author{E.~V.~Tkalya}
\email{tkalya@srd.sinp.msu.ru}

\affiliation{Skobeltsyn Institute of Nuclear Physics Lomonosov
Moscow State University, Leninskie gory, Moscow 119991, Russia}
\affiliation{Nuclear Safety Institute of RAS, Bol'shaya Tulskaya
52, Moscow 115191, Russia} \affiliation{National Research Nuclear
University MEPhI, 115409, Kashirskoe shosse 31, Moscow, Russia}

\date{\today}

\begin{abstract}
The magnetic hyperfine (MHF) splitting of the ground and
low-energy $3/2^+(7.8 \pm 0.5$~eV) levels in the $^{229}$Th
nucleus in muonic atom ($\mu^-_{1S_{1/2}}{}^{229}$Th$)^*$ has been
calculated considering the distribution of the nuclear
magnetization in the framework of collective nuclear model with
the wave functions of the Nilsson model for the unpaired neutron.
It is shown that (a) the deviation of MHF structure of the
isomeric state exceeds 100\% from its value for a point-like
nuclear magnetic dipole (the order of sublevels is reversed), (b)
partial inversion of levels of the $^{229}$Th ground-state doublet
and spontaneous decay of the ground state to the isomeric state
takes place, (c) the $E0$ transition which is sensitive to the
differences in the mean-square charge radii of the doublet states
is possible between the mixed sublevels with $F=2$, (d) the MHF
splitting of the $3/2^+$ isomeric state may be in the optical
range for certain values of the intrinsic $g_K$ factor and reduced
probability of the nuclear transition between the isomeric and
ground states.
\end{abstract}

\maketitle

The unique transition between the low-lying isomeric level
$3/2^+(E_{is} = 7.8\pm0.5$ eV) (its energy is measured in
\cite{Beck-07-s} and its existence is confirmed in
\cite{Wense-16-s}) and the ground $5/2^+(0.0)$ state in the
$^{229}$Th nucleus draws attention of specialists from different
areas of physics. The reason is the anomalous low energy of the
transition. Its proximity to the optical range gives us a hope for
a number of scientific breakthroughs that could have a significant
impact on technological development and applications. This is a
new metrological standard for time
\cite{Peik-03,Campbell-12-s,Kazakov-12-s} and a laser at nuclear
transition in the VUV range \cite{Tkalya-11}. The relative effect
of the variation of the fine structure constant $e^2$ (we use the
system of units $\hbar=c=1$) and the strong interaction parameter
$m_q/\Lambda_{QCD}$ \cite{Flambaum-06} are also of considerable
scientific interest. Finally, we mention the decay of the isomeric
nuclear level via the electronic bridge \cite{Strizhov-91}, high
sensitivity of the nuclear transition to the chemical environment
and the ability to use thorium isomer as a probe to study the
physicochemical properties of solids \cite{Strizhov-91}, the
cooperative spontaneous emission Dicke \cite{Dicke-54} in the
system of excited nuclei $^{229}$Th, accelerated $\alpha$-decay of
the $^{229}$Th nucleus via the isomeric state
\cite{Tkalya-00-PRC}. The behavior of the excited $^{229}$Th
nucleus inside dielectrics with a large band gap is of particular
interest \cite{Tkalya-00-JETPL}. Since there is no conversion
decay channel in such dielectric, the nucleus can absorb and emit
the VUV range photons directly, without interaction with the
electron shell \cite{Tkalya-00-PRC}. As a result, studying of
isomeric state by the optical methods becomes possible
\cite{Kazakov-12-s,Rellergert-10-s,Stellmer-15,Jeet-15-s,Yamaguchi-15-s}.

In this work the $^{229}$Th ground-state doublet is investigated
in muonic atom ($\mu^-_{1S_{1/2}}{}^{229}$Th$)^*$. The muon on the
$1S_{1/2}$ atomic orbit creates a very strong magnetic field at
the nucleus \cite{Wheeler-49,Kim-71}. The interaction of this
field with the magnetic moments of nuclear states leads to a
magnetic hyperfine (MHF) splitting of nuclear levels (see for
example
\cite{Wu-69,Johnson-68,Engfer-68,Baader-68,Klemt-70,Link-74,Ruetschi-84,Wycech-93,Karpeshin-98-s,Measday-01,Beloy-14}
and references therein). We demonstrate here that the MHF
splitting has a number of non-trivial features in the case of
($\mu^-_{1S_{1/2}}{}^{229}$Th$)^*$: the partial inversion of
nuclear sublevels and spontaneous decay of the ground state
$5/2^+$ to the isomeric $3/2^+$ state, the anomaly deviation of
MHF structure of the isomeric state from its value for a
point-like nucleus, an important role of the dynamic effect of
finite nuclear size (or the penetration effect) in the states
mixing, the possible existence of the electric monopole transition
and optical transitions between the MHF sublevels, etc. This
situation is very unusual and looks promising in regard to
experimental research.

{\it{The Fermi contact interaction}}. Let us consider the system
($\mu^-_{1S_{1/2}}{}^{229}$Th$)^*$ which consist of the muon bound
on the $1S_{1/2}$ shell of muonic atom and the $^{229}$Th nucleus.
The muon in the $(1S_{1/2})^1$ state results in a strong magnetic
field in the center of the $^{229}$Th nucleus. The value of this
field is given by the formula for the Fermi contact interaction
\begin{equation}
{\bf{H}}_{\mu}=-\frac{16\pi}{3} \frac{m_e} {m_{\mu}} \mu_B
\frac{{\mbox{\boldmath $\sigma$}}}{2} |\psi_{\mu}(0)|^2,
\label{eq:Hmu}
\end{equation}
where $m_e$ and $m_{\mu}$ are the masses of the electron and muon,
respectively, $\mu_B=e/2m_e$ is the Bohr magneton,
${\mbox{\boldmath{$\sigma$}}}$ are the Pauli matrixes, and
$\psi_{\mu}(0)$ is the amplitude of the muon Dirac wave function
at the origin.

The amplitude $\psi_{\mu}(0)$ can be calculated numerically by
solving the Dirac equations for the radial parts of the large,
$g(x)$, and small, $f(x)$, components of $\psi_{\mu}(x)$:
\begin{equation}
\begin{array}{l}
xg'(x)-b(E+1-V(x))xf(x)=0, \nonumber \\
xf'(x)+2f(x)+b(E-1-V(x))xg(x)=0. \nonumber
\end{array}
\label{eq:Dirac}
\end{equation}
Here $x=r/R_0$, where $r$ is the muon coordinate in the spherical
coordinate system, and $R_0=1.2A^{1/3}$ fm is the average radius
of the $^{229}$Th nucleus that has a form of charged sphere,
$b=m_{\mu}R_0$, $E$ and $V(x)$ are, respectively, the muon binding
and potential energies (in the units of $m_{\mu}$) in the field
produced by the nucleus protons. (For the lower muonic states,
electron screening plays a negligible role
\cite{Wu-69,Measday-01}. Therefore we neglect here the effects due
to the influence of the electron shell on the muon wave function.)

\begin{figure}
 \includegraphics[angle=0,width=0.8\hsize,keepaspectratio]{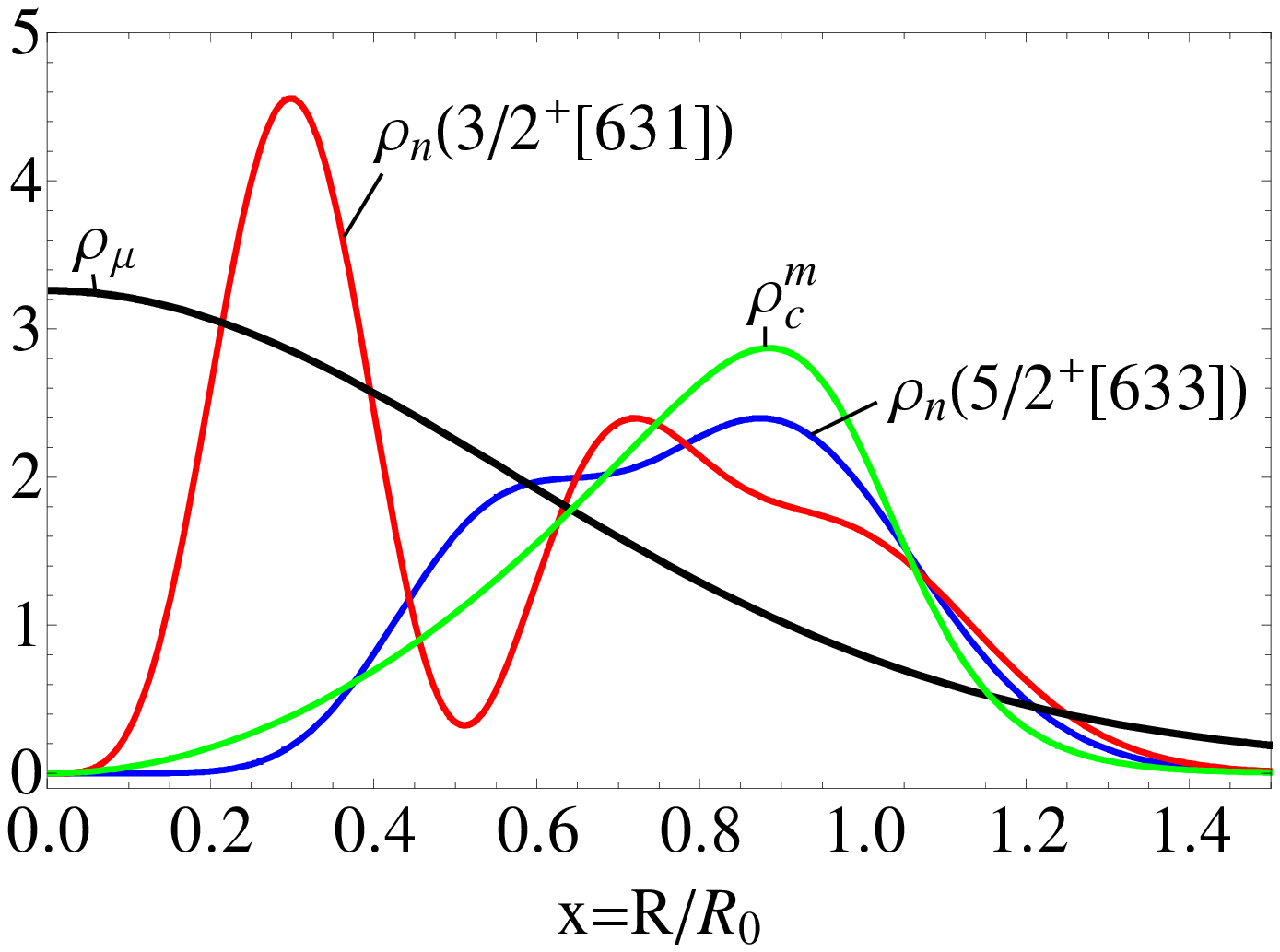}
  \caption{(color online). Dimensionless densities of muon ($\rho_{\mu}$),
  unpaired neutron ($\rho_n$) in the ground $5/2^+[633]$ state
  and isomeric $3/2^+[631]$ state, $\rho^m_{c}$ is the core magnetization.}
  \label{fig:Densities}
\end{figure}

We assume that the proton density of the nucleus has the Fermi
shape
$
\rho_p(x)=\rho_0/[1+exp((x-1)/\chi)],
$
where $\chi=[0.449+0.071(Z/N)]/R_0$ is diffuseness or the
half-density parameter of the proton density Fermi distribution
\cite{Seif-15}. The density is normalized by the condition
$\int_0^{\infty}\rho_p(x)x^2dx=Ze$, where $Z$ is the nucleus
charge. The muon wave function is normalized by the condition
$\int_0^{\infty}\rho_{\mu}(x)x^2dx=1$, where $\rho_{\mu}(x)$ is
the muon density $\rho_{\mu}(x)=g^2(x)+f^2(x)$. The result of
calculation of the muon density is presented in
Fig.~\ref{fig:Densities}. To evaluate the magnetic field one can
use Eq.~(\ref{eq:Hmu}) with the value of the muon wave function
given by
$
\psi_{\mu}(0)=Y_{00}(\vartheta,\varphi)g(0)/R_0^{3/2}
$,
where $Y_{00}(\vartheta,\varphi)$ is the spherical harmonic, and
from calculations it follows that $g(0) = \sqrt{\rho_{\mu}(0)} =
1.76$.

%
%

\begin{figure}
 \includegraphics[angle=0,width=0.98\hsize,keepaspectratio]
 {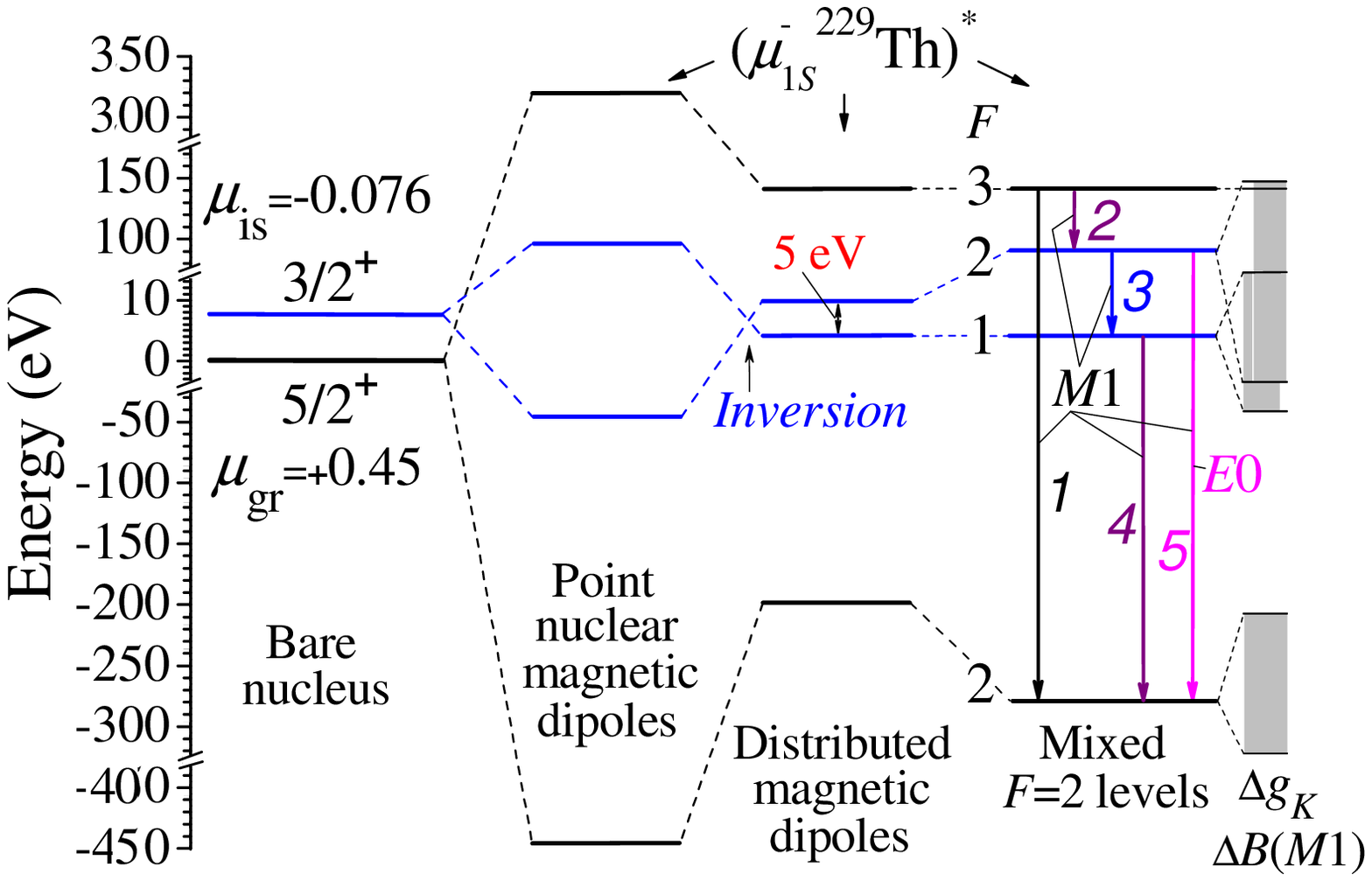}
  \caption{(color online). Magnetic hyperfine structure
  of the $^{229}$Th ground-state doublet in muonic atom in various models.
 The uncertainty range for the energy of the states due to variations
 of parameters $g_K$ and $B_{W.u.}(M1;3/2^+\rightarrow{}5/2^+)$ (see text for details)
 is shown on the right.}
  \label{fig:SchemeHFS}
\end{figure}

Thus, according to Eq.~(\ref{eq:Hmu}) the magnetic field at the
center of the $^{229}$Th nucleus is about 23 GT. Interaction of
point magnetic moments of the ground state ($\mu_{gr}=0.45$) and
isomeric state ($\mu_{is}=-0.076$) with the magnetic field leads
to a splitting of the nuclear levels. The energy of the sublevels
is determined by the formula
\begin{equation}
E=E_{int}\frac{F(F+1)-I(I+1)-s(s+1)}{2Is},
\label{eq:E}
\end{equation}
where $E_{int}=-\mu_{gr(is)}\mu_N{}H_{\mu}$ is the interaction
energy, $\mu_N=e/2M_p$ is the nuclear magneton ($M_p$ is the
proton mass), $I$ is the nuclear state spin, $s$ is the muon spin.
The quantum number $F$ takes two values $F=I\pm1/2$ for the ground
and isomeric states and determines the sublevels energy. The
resulting energy values are given in Fig.~\ref{fig:SchemeHFS}.

The MHF splitting found in the model of the Fermi contact
interaction is very significant. However, since the muon density
decreases quickly to the nuclear edge the obtained values are
grossly overestimated.

{\it{The distributed magnetic dipole model}}. The influence of the
finite nuclear size on the MHF splitting was first considered by
Bohr and Weisskopf \cite{Bohr-50}. Later the effect of the
distribution of nuclear magnetization on MHF structure in muonic
atoms was studied by Le Bellac \cite{LeBellac-63}. According to
their works, in the case of deformed nucleus the energy of
sublevels is given by Eq.~(\ref{eq:E}), where
\begin{equation}
E_{int}=\int{}d^3r\,{\bf{j}}({\bf{r}}) {\bf{A}}({\bf{r}})
\label{eq:Eint}
\end{equation}
is the energy of interaction of the muon current
${\bf{j}}({\bf{r}}) = -e\psi_{\mu}^+({\bf{r}}){\mbox{\boldmath
$\alpha$}} \psi_{\mu}({\bf{r}})$ (${\mbox{\boldmath
$\alpha$}}=\gamma^0{\mbox{\boldmath $\gamma$}}$, $\gamma$ are the
Dirac matrices) with the vector potential of the electromagnetic
field ${\bf{A}}({\bf{r}})$ generated by the magnetic moment of the
nucleus. For a system of ``rotating deformed core (with the
collective rotating angular momentum ${\mbox{\boldmath $\Re$}}$) +
unpaired neutron (with the spin ${\bf{S}}_n$)'', vector-potential
is determined by the relation \cite{Bohr-50,LeBellac-63}
\begin{equation}
{\bf{A}}({\bf{r}}) =
-\int{}d^3R\left[\rho_n({\bf{R}})g_S{\bf{S}}_n +
\rho_{c}^m({\bf{R}})g_R{\mbox{\boldmath $\Re$}} \right] \times
{\mbox{\boldmath $\nabla$}}_r\frac{1}{|{\bf{r}}-{\bf{R}}|},
 \label{eq:A}
\end{equation}
where $\rho_n({\bf{R}})$ is the distribution of the spin part of
the nuclear moment and $\rho_{c}^m({\bf{R}})$ is the distribution
of the core magnetization, $g_S$ is the spin $g$-factor, and $g_R$
is the core gyromagnetic ratio. The distributions
$\rho_n({\bf{R}})$ and $\rho_{c}^m({\bf{R}})$ are normalized:
$\int{}d^3R \rho_n({\bf{R}})=1$, $\int{}d^3R
\rho_{c}^m({\bf{R}})=1$.

Here we use the standard nuclear wave function \cite{Bohr-98-II}
$
\Psi_{MK}^I=\sqrt{(2I+1)/8\pi^2}D_{MK}^I({\bf{\Omega}})\varphi_K({\bf{R}})
$,
where $D_{MK}^I({\bf{\Omega}})$ is the Wigner $D$-function of the
Euler angles are denoted, collectively, by ${\bf{\Omega}}$,
$\varphi_K({\bf{R}})$ is the wave function of external neutron
coupled to the core, $K$ is the component of $I$ along the
symmetry axis of the nucleus, and $M$ is the component of $I$
along the direction of magnetic field.

As follows from Eqs.~(\ref{eq:Eint}--\ref{eq:A}), $E_{int}$
consists of two parts. The first part is the interaction of the
muon with the external unpaired neutron and the second one is the
interaction of the muon with the rotating charged nuclear core.
These energies are calculated in accordance with formulas from
\cite{LeBellac-63}. In our case of the muon interacts with the
nucleus in the head levels of rotational bands (for such states we
have $K=I$), and two contributions take the following form:
\begin{eqnarray}
 E_{int}^{n \choose{}core} &=& E_{0}\frac{I}{I+1}
{I g_K \choose{} g_R} \left\{ \langle{\cal{M}}\rangle - \int{}
{\rho_{n}({\bf{y}}) \choose{} \rho_{c}^m({\bf{y}})} d^3y
 \right.
         \nonumber\\
  &\times& \left. \int_0^y\left[1-x^3 {\Theta(I,\theta) \choose
1} \right]f(x)g(x)dx\right\}. \label{eq:EintNeutronCore}
\end{eqnarray}
Here, $E_{0}=-2e^2M_p/[3(M_pR_0)^2]$, $g_K$ is the intrinsic $g$
factor,
$\rho_{n}({\bf{y}})=\varphi_K({\bf{y}})^*\varphi_K({\bf{y}})$,
${\bf{y}}={\bf{R}}/R_0$,
$\Theta(I,\theta)=\sqrt{4\pi/5}Y_{20}(\theta)(2I+1)/[I(2I+3)]$.
The first term in the square brackets in
Eq.~(\ref{eq:EintNeutronCore}), $\langle{\cal{M}}\rangle =
\int_0^{\infty}f(x)g(x)dx = -0.1895$, corresponds to the
interaction of the muon with a point nuclear magnetic dipole. The
resulting energy sublevels are close to the values calculated with
Eq.~(\ref{eq:Hmu}).

For the unpaired neutron the wave functions $\varphi_K$ were taken
from the Nilsson model. The structure of the intrinsic state
$\varphi_K$ of the $^{229}$Th ground state $5/2^+(0.0)$ is
$K^{\pi}[Nn_z\Lambda]=5/2^+[633]$. The structure of the isomeric
state $3/2^+$(7.8 eV) is $3/2^+[631]$ \cite{Helmer-94}. For each
of these states, the wave function has the form
$ \varphi_K
=\phi_{\Lambda}(\varphi)\phi_{\Lambda,n_r}(\eta)\phi_{n_z}(\zeta)
$
where the quantum number $n_r=(N-n_z-\Lambda)/2$, the variables on
the axes $\zeta=R_0\sqrt{M_p\omega_z}y$cos$\theta$,
$\eta=R_0\sqrt{M_p\omega_{\perp}}y$sin$\theta$, the energies of
the oscillatory quanta $\omega_z=\omega_0\sqrt{1+2\delta/3}$ and
$\omega_{\perp}=\omega_0\sqrt{1-4\delta/3}$, where
$\omega_0=41/A^{1/3}$ MeV is the harmonic oscillator frequency,
$\delta=0.95\beta$, and $\beta$ is the parameter of the
deformation of the nucleus defined in terms of the expansion of
the radius parameter $R=R_0(1+\beta{}Y_{20}(\theta)+\ldots)$.

The constituent wave functions are as follows:
$\phi_{\Lambda}(\varphi) = e^{i\Lambda\varphi}/\sqrt{2\pi}$,
$\phi_{\Lambda,n_r}(\eta) = e^{-\eta^2/2} \eta^{\Lambda}
L_{n_r}^{(\Lambda)}(\eta^2)/N_{\eta}$,
$\phi_{n_z}(\zeta) = e^{-\zeta^2/2}H_{n_z}(\zeta)/N_{\zeta}$,
where $L_{n_r}^{(\Lambda)}$ is the generalized Laguerre
polynomial, $H_{n_z}$ is the Hermite polynomial
\cite{Abramowitz-64}, $N_{\eta,\zeta}$ are the normalization
factors. The density distributions of the unpaired neutron in the
states $5/2^+[633]$ and $3/2^+[631]$ averaged over the angles
$\theta$ and $\varphi$ are shown in Fig.~\ref{fig:Densities}. In
our numerical calculations we took into account the asymmetry of
the nucleon wave functions in Eq.~(\ref{eq:EintNeutronCore}), but
neglected the small difference between $\omega_z$ and
$\omega_{\perp}$.

For the core magnetization we used the classical density of
magnetic moment, $\rho^m_{c}\propto{} x^2/[(1+\exp((x-1)/\chi)]$,
obtained from proton density $\rho_p$ by averaging over the
angles. Such quadratic dependence was used in
\cite{Johnson-68,Finkbeiner-93}. The normalized function
$\rho^m_{c}$ is shown in Fig.~\ref{fig:Densities}.

The resulting scheme of MHF splitting for
($\mu^-_{1S_{1/2}}{}^{229}$Th$)^*$ is shown in
Fig.~\ref{fig:SchemeHFS}. For $g$-factors of the ground state we
have used values, which are accepted nowadays: $g_R = 0.309$, $g_K
= 0.128$ \cite{Bemis-88-s}. The reduction of MHF structure in
comparison with the model of point nuclear magnetic dipole is
about 56\% for the $5/2^+(0.0)$ state.

%
%

For calculation of the isomeric state we have taken $g_R=0.309$
and $g_K=-0.29$ which is obtained from the mean value
$|g_K-g_R|=0.60$ (the values $|g_K-g_R|=0.59\pm0.14$ and
$0.61\pm0.10$ were measured in \cite{Kroger-76}). The gyromagnetic
ratio $g_R=0.309\pm 0.016$ from \cite{Bemis-88-s} is determined
with a much higher precision than $|g_K-g_R|$ for the band
$3/2^+[631]$, and existing uncertainty in $|g_K-g_R|$ is related
exclusively with $g_K$: $g_K=0.29 \pm 0.12$. This leads to
uncertainty in the position of levels (see in
Fig.~\ref{fig:SchemeHFS}).

\begin{figure}
 \includegraphics[angle=0,width=1.0\hsize,keepaspectratio]
 {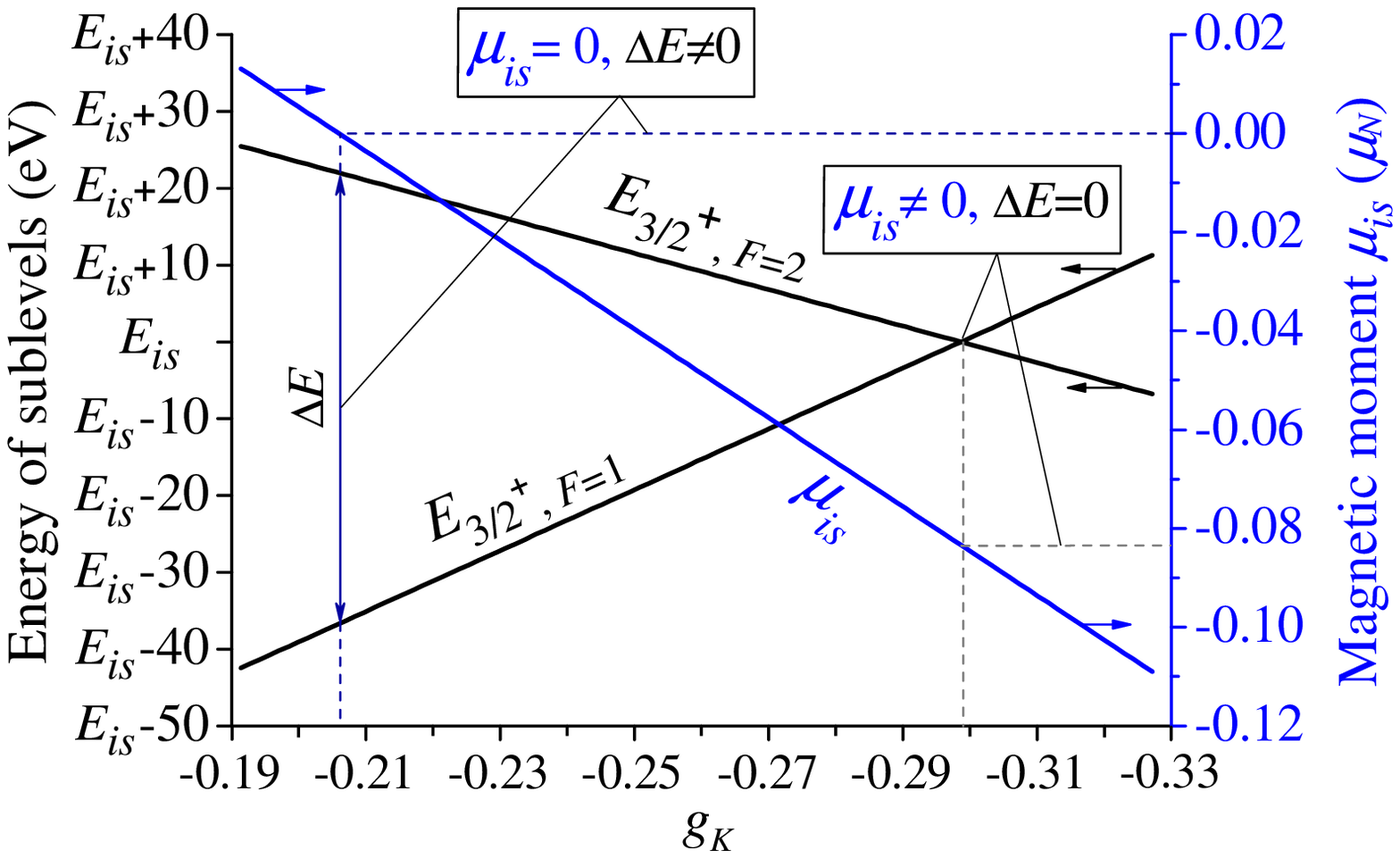}
  \caption{(color online). Imbalance of the MHF interaction for the
  composed magnetic moment of the isomeric state in $^{229}$Th:
  the energies of the sublevels relative to $E_{is}$ and the magnetic moment $\mu_{is}$
  as a function of the gyromagnetic factor $g_K$ in the absence
  of mixing of the states with $F=2$ (see text for details).}
  \label{fig:EgK}
\end{figure}

A somewhat paradoxical situation can take place because of the
complex structure of the magnetic moment of the isomeric state and
the behavior of the muon wave function (currently we consider a
variant without mixing of the states with the equal values of
$F$). From Fig.~\ref{fig:EgK} it follows that in the range $-0.30
<g_K <-0.29$ the $3/2^+[631]$ state has a nonzero magnetic moment,
whereas the MHF splitting is absent or very weak. Conversely, the
magnetic moment of the isomeric level equals to zero for
$g_K\approx-0.206$, while the MHF splitting is relatively large.
The reason is the following. The magnetic field generated by the
spin of the nucleon is sensitive to the non-sphericity of the wave
functions $\varphi_K$.  This leads to the appearance of the
additional factor $\Theta(I,\theta)$ in the spin part of the
Eq.~(\ref{eq:EintNeutronCore}) \cite{Bohr-50,LeBellac-63}.
Averaging over the angles reduces the spin contribution in respect
to the orbital part. A small imbalance emerged in the system leads
to the violation of the ``fine tuning'' between the spin and
orbital parts of the magnetic moment and to the effect described
above. This mechanism can also occur in other nuclei with low
energy (up to some kiloelectronvolts) levels.

%
%

{\it{Mixing of the sublevels with $F=2$}}. To find the final
position of the sublevels we now consider the mixing of the states
with $F=2$ \cite{Wycech-93}. The interaction energy, $\cal{E}$, of
the nuclear and muon currents during the transition between the $
|3/2^+,F = 2\rangle$ sublevel with the energy $E_1$ and the
$|5/2^+,F=2\rangle$ sublevel with the energy $E_2$ can be found
from equations given in Refs. \cite{Tkalya-92,Tkalya-94}. They
generalize the static Bohr-Weisskopf effect for the case of
nuclear excitation at the electron (muon) transitions in the
atomic shell. For $M1$ transition we obtain
$$
{\cal{E}} = E_0 \xi \langle{\cal{M}}\rangle{}
\sqrt{(15/2)B_{W.u.}(M1;3/2^+\rightarrow{}5/2^+)},
$$
where $B_{W.u.}(M1;3/2^+\rightarrow{}5/2^+) = 3.0\times10^{-2}$ is
the reduced probability of the nuclear isomeric transition in
Weisskopf's units \cite{Tkalya-15}, $\xi$ is a factor that takes
into account the dynamic effect of the nuclear size
\cite{Tkalya-94} or the penetration effect \cite{Church-60}.
Calculation of the nuclear current with the neutron wave function
in the Nilsson model gives the value of $\xi = 0.45$. As a result,
we have ${\cal{E}} \simeq 150$ eV.

The energies of the new sublevels with $F=2$ are calculated
according to the formulas \cite{Davydov-65}:
$$
E_{1',2'}=[E_1+E_2 \pm \sqrt{(E_1-E_2)^2+(2{\cal{E}})^2}]/2,
$$
where $E_{1'(2')}$ are the energies of the new sublevels
$|3/2^+(5/2^+),F=2\rangle{}'$. We emphasize that these energies
are valid for the most probable values of $g_K$ and
$B_{W.u.}(M1;3/2^+\rightarrow{}5/2^+)$. Variations of the
parameter $g_K$ in the range ($g_K=0.29 \pm 0.12$) and the reduced
probability of the nuclear transition (currently
$3\times10^{-3}\leq{}B_{W.u.}(M1;3/2^+\rightarrow{}5/2^+)\leq5\times10^{-2}$
\cite{Tkalya-15}) gives a fairly large area of uncertainty (see in
Fig.~\ref{fig:SchemeHFS}) in the position of the levels.

\begin{figure}
 \includegraphics[angle=0,width=0.6\hsize,keepaspectratio]{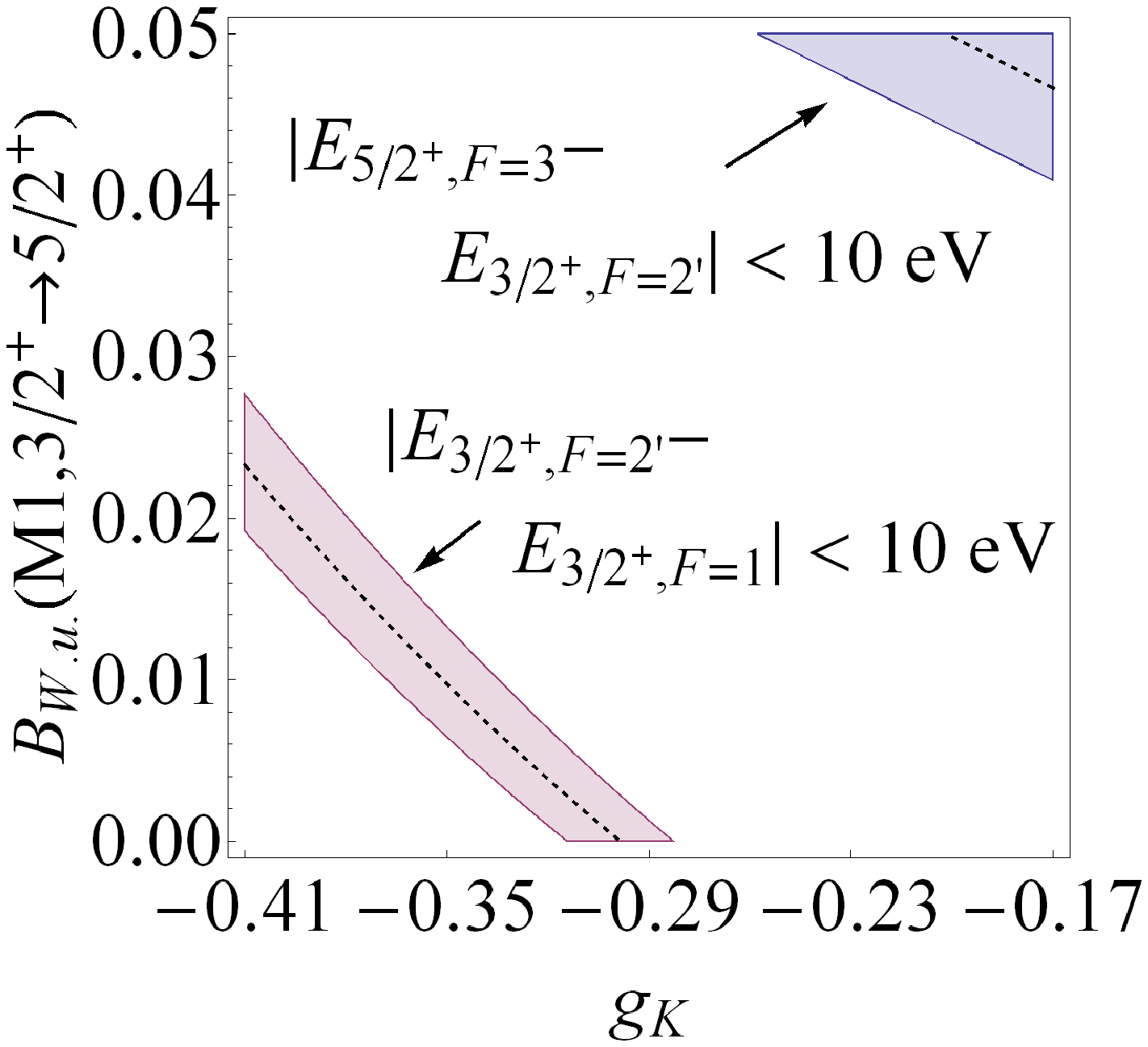}
  \caption{(color online). Range of values of $g_K$ and
  $B_{W.u.}(M1;3/2^+\rightarrow{}5/2^+)$ at which the transitions between the
  sublevels lie in the optical or VUV ranges. The dotted lines show the
  areas where the sublevels have the same energy.}
  \label{fig:OpticalRange-gB}
\end{figure}

In Fig.~\ref{fig:OpticalRange-gB} we reproduce values of $g_K$ and
$B_{W.u.}(M1;3/2^+\rightarrow{}5/2^+)$ with the energy difference
between the sublevels less than 10 eV. The existing of the optical
range for the transitions $|5/2^+,F=3\rangle \rightarrow
|3/2^+,F=2\rangle{}'$ and $|3/2^+,F=2\rangle{}' \rightarrow
|3/2^+,F=1\rangle$ is an unusual feature of the MHF structure in
$(\mu^-_{1S_{1/2}}{}^{229}$Th$)^*$. It gives a hope that advanced
optical methods can be applied for the study of this extraordinary
nuclear state.

%
%

{\it{Transitions between sublevels}}. Both sublevels of the
isomeric state $3/2^+(7.8$ eV) lie below the ground-state sublevel
$|5/2^+,F=3\rangle$. As a result, spontaneous transitions to the
isomeric level accompanied by its population become possible.

Mixing of the sublevels with $F=2$ significantly increases the
probability of the transitions 2 and 4 in Fig.~\ref{fig:SchemeHFS}
between the sublevels of the ground and isomeric states. The wave
functions of the new sublevels have the form
\begin{equation}
\begin{array}{l}
|3/2^+,F=2\rangle{}' = \sqrt{1-b^2}|3/2^+,F=2\rangle + b|5/2^+,F=2\rangle \nonumber \\
|5/2^+,F=2\rangle{}' = - b|3/2^+,F=2\rangle{} +
\sqrt{1-b^2}|5/2^+,F=2\rangle{} ,\nonumber
\end{array}
\end{equation}
where $b=(E_{1'}-E_1)/\sqrt{(E_{1'}-E_1)^2+{\cal{E}}^2}\simeq
0.47$ \cite{Davydov-65}. Accordingly, the component of the
transition, which connects the state $|5/2^+,F=3\rangle$ with
$b|5/2^+,F=2\rangle$ gives the main contribution to the transition
2 in Fig.~\ref{fig:SchemeHFS}. This transition occurs via a spin
flip of the muon without changing nuclear state.

The main decay channels of the $|5/2^+,F=3\rangle$ sublevel is the
transition to the $|5/2^+,F=2\rangle$ ground state sublevel
(labeled as 1 in Fig.~\ref{fig:SchemeHFS}). The probability of the
transition 1 calculated by means of formulas of Refs.
\cite{Winston-63,Karpeshin-98-s} is $2.8\times10^{-11}$ eV. The
transition is accompanied by the emission of conversion electrons.
Muon in ($\mu^-_{1S_{1/2}}{}^{229}$Th$)^*$ is practically inside
the thorium nucleus. Electronic shell perceives the system ``muon
+ Thorium nucleus'' as the Actinium nucleus of charge 89.
Therefore, the internal conversion will take place in the electron
shell of the Ac atom. For the transition 1 the internal conversion
coefficient $\alpha_{M1}$ is equal to $6.6\times10^5$ (it have
been found using the code described in \cite{Strizhov-91}) with
the full width $\Gamma_{tot}=1.8\times10^{-5}$ eV. This means that
the half-life of the sublevel $|5/2^+,F=3\rangle$ is less than
$2.5\times10^{-11}$ s. I.e. the relaxation of this level is
completed prior to the muon absorption ($\sim10^{-7}$ s) or the
muon decay ($2\times10^{-6}$ s).

Taking into account the coefficient $b^2$, the radiation width of
the transition 2 is $1.1\times10^{-14}$ eV and the total width
equals to $7.0\times10^{-7}$ eV ($\alpha_{M1}=6.0\times10^7$).
Thus, the probability of the isomeric state excitation at the
decay of the ground state is 3-4\%. Modern muon factories generate
of $10^5$ muonic atoms per second. Thus we can expect the
formation of the order of $N_{is}\simeq3\times10^3$ isomeric
nuclei per second. From the measurements of the corresponding
conversion electrons one can hope to identify experimentally the
fast transitions 3, 4, and 5. They are comparable in intensity
with the transitions 1 and 2. The measurement of the parameters of
the transitions can give information about $g_K$ and
$B(M1;3/2^+\rightarrow{}5/2^+)$.

The value $N_{is}\simeq3\times10^3$ s$^{-1}$ is a lower estimate.
The muon capture by atom is followed by a cascade of muon
transitions in the atomic shell. The process of nonradiative
nuclear excitation by means of direct energy transfer from the
excited atomic shell to the nucleus via the virtual $X$-photon is
possible if the muon transition is close in energy and coincides
in type with the nuclear one (see for example \cite{Ruetschi-84}).
This effect was predicted by Wheeler \cite{Wheeler-49}. In the
case of resonant excitation of the levels of the $5/2^+[633]$
rotational band the probability of the population of the
$3/2^+[631]$ isomeric state is estimated by 1-2\%. (This value
corresponds to the probability of the isomer population at the
$\alpha$ decay of $^{233}$U, which involves mainly the levels of
the $5/2^+[633]$ band in $^{229}$Th.) However, a precise account
of the isomer population in muonic transitions can be given only
experimentally.

Another interesting consequence of the $F=2$ states mixing is the
possible existence of the $E0$ component at the transition 5 in
Fig.~\ref{fig:SchemeHFS}. The $E0$ transition is sensitive to the
differences in the mean-square charge radii
$\langle{}R_p^2\rangle$ \cite{Church-56}. The probability of the
transition depends on the $E0$ transition strengths $\rho(E0)^2$,
which is proportional to
$b^2(1-b^2)(\langle{}R_p^2\rangle{}_{5/2^+}-
\langle{}R_p^2\rangle{}_{3/2^+})^2/R_0^4$. $\rho(E0)^2=0$ in the
framework of the simplified model for the charge distribution
$\rho_p$ used in this work. In reality the radii
$\langle{}R_p^2\rangle{}_{3/2^+}$ and
$\langle{}R_p^2\rangle{}_{5/2^+}$ can differ in magnitude and the
detection of the $E0$ transition would be a step towards a better
understanding of the properties of the low-energy doublet in
$^{229}$Th.

This research was carried out by a grant of Russian Science
Foundation (project No 16-12-00001).

\bibstyle{apsrev}


\begin{thebibliography}{50}
\expandafter\ifx\csname
natexlab\endcsname\relax\def\natexlab#1{#1}\fi
\expandafter\ifx\csname bibnamefont\endcsname\relax
  \def\bibnamefont#1{#1}\fi
\expandafter\ifx\csname bibfnamefont\endcsname\relax
  \def\bibfnamefont#1{#1}\fi
\expandafter\ifx\csname citenamefont\endcsname\relax
  \def\citenamefont#1{#1}\fi
\expandafter\ifx\csname url\endcsname\relax
  \def\url#1{\texttt{#1}}\fi
\expandafter\ifx\csname
urlprefix\endcsname\relax\def\urlprefix{URL }\fi
\providecommand{\bibinfo}[2]{#2}
\providecommand{\eprint}[2][]{\url{#2}}

\bibitem[{\citenamefont{Beck et~al.}(2007)\citenamefont{Beck, Becker,
  Beiersdorfer et~al.}}]{Beck-07-s}
\bibinfo{author}{\bibfnamefont{B.~R.} \bibnamefont{Beck}},
  \bibinfo{author}{\bibfnamefont{J.~A.} \bibnamefont{Becker}},
  \bibinfo{author}{\bibfnamefont{P.}~\bibnamefont{Beiersdorfer}},
  \bibnamefont{et~al.}, \bibinfo{journal}{Phys. Rev. Lett.}
  \textbf{\bibinfo{volume}{98}}, \bibinfo{pages}{142501}
  (\bibinfo{year}{2007}).

\bibitem[{\citenamefont{Wense et~al.}(2016)\citenamefont{Wense, Seiferle,
  Laatiaoui et~al.}}]{Wense-16-s}
\bibinfo{author}{\bibfnamefont{L.~von~der} \bibnamefont{Wense}},
  \bibinfo{author}{\bibfnamefont{B.} \bibnamefont{Seiferle}},
  \bibinfo{author}{\bibfnamefont{M.}~\bibnamefont{Laatiaoui}},
  \bibnamefont{et~al.}, \bibinfo{journal}{Nature}
  \textbf{\bibinfo{volume}{533}}, \bibinfo{pages}{47}
  (\bibinfo{year}{2016}).

\bibitem[{\citenamefont{Peik and Tamm}(2000)}]{Peik-03}
\bibinfo{author}{\bibfnamefont{E.}~\bibnamefont{Peik}} \bibnamefont{and}
  \bibinfo{author}{\bibfnamefont{C.}~\bibnamefont{Tamm}},
  \bibinfo{journal}{Europhys. Lett.} \textbf{\bibinfo{volume}{61}},
  \bibinfo{pages}{181} (\bibinfo{year}{2000}).

\bibitem[{\citenamefont{Campbell et~al.}(2012)\citenamefont{Campbell, Radnaev,
  Kuzmich et~al.}}]{Campbell-12-s}
\bibinfo{author}{\bibfnamefont{C.~J.} \bibnamefont{Campbell}},
  \bibinfo{author}{\bibfnamefont{A.~G.} \bibnamefont{Radnaev}},
  \bibinfo{author}{\bibfnamefont{A.}~\bibnamefont{Kuzmich}},
  \bibnamefont{et~al.}, \bibinfo{journal}{Phys. Rev. Lett.}
  \textbf{\bibinfo{volume}{108}}, \bibinfo{pages}{120802}
  (\bibinfo{year}{2012}).

\bibitem[{\citenamefont{Kazakov et~al.}(2012)\citenamefont{Kazakov, Litvinov,
  Romanenko et~al.}}]{Kazakov-12-s}
\bibinfo{author}{\bibfnamefont{G.~A.} \bibnamefont{Kazakov}},
  \bibinfo{author}{\bibfnamefont{A.~N.} \bibnamefont{Litvinov}},
  \bibinfo{author}{\bibfnamefont{V.~I.} \bibnamefont{Romanenko}},
  \bibnamefont{et~al.}, \bibinfo{journal}{New J. Phys.}
  \textbf{\bibinfo{volume}{14}}, \bibinfo{pages}{083019}
  (\bibinfo{year}{2012}).

\bibitem[{\citenamefont{Tkalya}(2011)}]{Tkalya-11}
\bibinfo{author}{\bibfnamefont{E.~V.} \bibnamefont{Tkalya}},
  \bibinfo{journal}{Phys. Rev. Lett.} \textbf{\bibinfo{volume}{106}},
  \bibinfo{pages}{162501} (\bibinfo{year}{2011}).

\bibitem[{\citenamefont{Flambaum}(2006)}]{Flambaum-06}
\bibinfo{author}{\bibfnamefont{V.~V.} \bibnamefont{Flambaum}},
  \bibinfo{journal}{Phys. Rev. Lett.} \textbf{\bibinfo{volume}{97}},
  \bibinfo{pages}{092502} (\bibinfo{year}{2006}).

\bibitem[{\citenamefont{Strizhov and Tkalya}(1991)}]{Strizhov-91}
\bibinfo{author}{\bibfnamefont{V.~F.} \bibnamefont{Strizhov}} \bibnamefont{and}
  \bibinfo{author}{\bibfnamefont{E.~V.} \bibnamefont{Tkalya}},
  \bibinfo{journal}{Sov. Phys. JETP} \textbf{\bibinfo{volume}{72}},
  \bibinfo{pages}{387} (\bibinfo{year}{1991}).

\bibitem[{\citenamefont{Dicke}(1954)}]{Dicke-54}
\bibinfo{author}{\bibfnamefont{R.~H.} \bibnamefont{Dicke}},
  \bibinfo{journal}{Phys. Rev.} \textbf{\bibinfo{volume}{93}},
  \bibinfo{pages}{99} (\bibinfo{year}{1954}).

\bibitem[{\citenamefont{Tkalya et~al.}(2000)\citenamefont{Tkalya, Zherikhin,
  and Zhudov}}]{Tkalya-00-PRC}
\bibinfo{author}{\bibfnamefont{E.~V.} \bibnamefont{Tkalya}},
  \bibinfo{author}{\bibfnamefont{A.~N.} \bibnamefont{Zherikhin}},
  \bibnamefont{and} \bibinfo{author}{\bibfnamefont{V.~I.}
  \bibnamefont{Zhudov}}, \bibinfo{journal}{Phys. Rev. C}
  \textbf{\bibinfo{volume}{61}}, \bibinfo{pages}{064308}
  (\bibinfo{year}{2000}).

\bibitem[{\citenamefont{Tkalya}(2000)}]{Tkalya-00-JETPL}
\bibinfo{author}{\bibfnamefont{E.~V.} \bibnamefont{Tkalya}},
  \bibinfo{journal}{JETP Lett.} \textbf{\bibinfo{volume}{71}},
  \bibinfo{pages}{311} (\bibinfo{year}{2000}).

\bibitem[{\citenamefont{Rellergert et~al.}(2010)\citenamefont{Rellergert,
  DeMille, Greco et~al.}}]{Rellergert-10-s}
\bibinfo{author}{\bibfnamefont{W.~G.} \bibnamefont{Rellergert}},
  \bibinfo{author}{\bibfnamefont{D.}~\bibnamefont{DeMille}},
  \bibinfo{author}{\bibfnamefont{R.~R.} \bibnamefont{Greco}},
  \bibnamefont{et~al.}, \bibinfo{journal}{Phys. Rev. Lett.}
  \textbf{\bibinfo{volume}{104}}, \bibinfo{pages}{200802}
  (\bibinfo{year}{2010}).

\bibitem[{\citenamefont{Stellmer et~al.}(2015)\citenamefont{Stellmer, Schreitl,
  and Schumm}}]{Stellmer-15}
\bibinfo{author}{\bibfnamefont{S.}~\bibnamefont{Stellmer}},
  \bibinfo{author}{\bibfnamefont{M.}~\bibnamefont{Schreitl}}, \bibnamefont{and}
  \bibinfo{author}{\bibfnamefont{T.}~\bibnamefont{Schumm}},
  \bibinfo{journal}{Sci. Rep.} \textbf{\bibinfo{volume}{5}},
  \bibinfo{pages}{15580} (\bibinfo{year}{2015}).

\bibitem[{\citenamefont{Jeet et~al.}(2015)\citenamefont{Jeet, Schneider,
  Sullivan et~al.}}]{Jeet-15-s}
\bibinfo{author}{\bibfnamefont{J.}~\bibnamefont{Jeet}},
  \bibinfo{author}{\bibfnamefont{C.}~\bibnamefont{Schneider}},
  \bibinfo{author}{\bibfnamefont{S.~T.} \bibnamefont{Sullivan}},
  \bibnamefont{et~al.}, \bibinfo{journal}{Phys. Rev. Lett.}
  \textbf{\bibinfo{volume}{114}}, \bibinfo{pages}{253001}
  (\bibinfo{year}{2015}).

\bibitem[{\citenamefont{Yamaguchi et~al.}(2015)\citenamefont{Yamaguchi, Kolbe,
  Kaser et~al.}}]{Yamaguchi-15-s}
\bibinfo{author}{\bibfnamefont{A.}~\bibnamefont{Yamaguchi}},
  \bibinfo{author}{\bibfnamefont{M.}~\bibnamefont{Kolbe}},
  \bibinfo{author}{\bibfnamefont{H.}~\bibnamefont{Kaser}},
  \bibnamefont{et~al.}, \bibinfo{journal}{New J. Phys.}
  \textbf{\bibinfo{volume}{17}}, \bibinfo{pages}{053053}
  (\bibinfo{year}{2015}).

\bibitem[{\citenamefont{Wheeler}(1949)}]{Wheeler-49}
\bibinfo{author}{\bibfnamefont{J.~A.} \bibnamefont{Wheeler}},
  \bibinfo{journal}{Rev. Mod. Phys.} \textbf{\bibinfo{volume}{21}},
  \bibinfo{pages}{133} (\bibinfo{year}{1949}).

\bibitem[{\citenamefont{Kim}(1971)}]{Kim-71}
\bibinfo{author}{\bibfnamefont{Y.~N.} \bibnamefont{Kim}},
  \emph{\bibinfo{title}{Mesic Atoms and Nuclear Structure}}
  (\bibinfo{publisher}{North-Holland Publ. Comp.},
  \bibinfo{address}{Amsterdam}, \bibinfo{year}{1971}).

\bibitem[{\citenamefont{Wu and Wilets}(1969)}]{Wu-69}
\bibinfo{author}{\bibfnamefont{C.~S.} \bibnamefont{Wu}} \bibnamefont{and}
  \bibinfo{author}{\bibfnamefont{L.}~\bibnamefont{Wilets}},
  \bibinfo{journal}{Annu. Rev. Nucl. Sci.} \textbf{\bibinfo{volume}{19}},
  \bibinfo{pages}{527} (\bibinfo{year}{1969}).

\bibitem[{\citenamefont{Johnson and Sorensen}(1968)}]{Johnson-68}
\bibinfo{author}{\bibfnamefont{J.}~\bibnamefont{Johnson}} \bibnamefont{and}
  \bibinfo{author}{\bibfnamefont{R.~A.} \bibnamefont{Sorensen}},
  \bibinfo{journal}{Phys. Lett. B} \textbf{\bibinfo{volume}{28}},
  \bibinfo{pages}{700} (\bibinfo{year}{1968}).

\bibitem[{\citenamefont{Engfer and Scheck}(1968)}]{Engfer-68}
\bibinfo{author}{\bibfnamefont{R.}~\bibnamefont{Engfer}} \bibnamefont{and}
  \bibinfo{author}{\bibfnamefont{F.}~\bibnamefont{Scheck}},
  \bibinfo{journal}{Z. Phys.} \textbf{\bibinfo{volume}{216}},
  \bibinfo{pages}{274} (\bibinfo{year}{1968}).

\bibitem[{\citenamefont{Baader et~al.}(1968)\citenamefont{Baader, Backe, Engfer
  et~al.}}]{Baader-68}
\bibinfo{author}{\bibfnamefont{R.}~\bibnamefont{Baader}},
  \bibinfo{author}{\bibfnamefont{H.}~\bibnamefont{Backe}},
  \bibinfo{author}{\bibfnamefont{R.}~\bibnamefont{Engfer}},
  \bibnamefont{et~al.}, \bibinfo{journal}{Phys. Lett. B}
  \textbf{\bibinfo{volume}{27}}, \bibinfo{pages}{428} (\bibinfo{year}{1968}).

\bibitem[{\citenamefont{Klemt et~al.}(1970)\citenamefont{Klemt, Speth, and
  Goke}}]{Klemt-70}
\bibinfo{author}{\bibfnamefont{V.}~\bibnamefont{Klemt}},
  \bibinfo{author}{\bibfnamefont{J.}~\bibnamefont{Speth}}, \bibnamefont{and}
  \bibinfo{author}{\bibfnamefont{K.}~\bibnamefont{Goke}},
  \bibinfo{journal}{Phys. Lett. B} \textbf{\bibinfo{volume}{33}},
  \bibinfo{pages}{331} (\bibinfo{year}{1970}).

\bibitem[{\citenamefont{Link}(1974)}]{Link-74}
\bibinfo{author}{\bibfnamefont{R.}~\bibnamefont{Link}}, \bibinfo{journal}{Z.
  Physik} \textbf{\bibinfo{volume}{269}}, \bibinfo{pages}{163}
  (\bibinfo{year}{1974}).

\bibitem[{\citenamefont{Ruetschi et~al.}(1984)\citenamefont{Ruetschi,
  Schellenberg, Phan et~al.}}]{Ruetschi-84}
\bibinfo{author}{\bibfnamefont{A.}~\bibnamefont{Ruetschi}},
  \bibinfo{author}{\bibfnamefont{L.}~\bibnamefont{Schellenberg}},
  \bibinfo{author}{\bibfnamefont{T.~Q.} \bibnamefont{Phan}},
  \bibnamefont{et~al.}, \bibinfo{journal}{Nucl. Phys. A}
  \textbf{\bibinfo{volume}{422}}, \bibinfo{pages}{461} (\bibinfo{year}{1984}).

\bibitem[{\citenamefont{Wycech and Zylicz}(1993)}]{Wycech-93}
\bibinfo{author}{\bibfnamefont{S.}~\bibnamefont{Wycech}} \bibnamefont{and}
  \bibinfo{author}{\bibfnamefont{J.}~\bibnamefont{Zylicz}},
  \bibinfo{journal}{Acta Phys. Pol. B} \textbf{\bibinfo{volume}{24}},
  \bibinfo{pages}{637} (\bibinfo{year}{1993}).

\bibitem[{\citenamefont{Karpeshin et~al.}(1998)\citenamefont{Karpeshin, Wycech,
  Band et~al.}}]{Karpeshin-98-s}
\bibinfo{author}{\bibfnamefont{F.~F.} \bibnamefont{Karpeshin}},
  \bibinfo{author}{\bibfnamefont{S.}~\bibnamefont{Wycech}},
  \bibinfo{author}{\bibfnamefont{I.~M.} \bibnamefont{Band}},
  \bibnamefont{et~al.}, \bibinfo{journal}{Phys. Rev. C}
  \textbf{\bibinfo{volume}{57}}, \bibinfo{pages}{3085} (\bibinfo{year}{1998}).

\bibitem[{\citenamefont{Measday}(2001)}]{Measday-01}
\bibinfo{author}{\bibfnamefont{D.~F.} \bibnamefont{Measday}},
  \bibinfo{journal}{Phys. Rep.} \textbf{\bibinfo{volume}{354}},
  \bibinfo{pages}{243} (\bibinfo{year}{2001}).

\bibitem[{\citenamefont{Beloy}(2014)}]{Beloy-14}
\bibinfo{author}{\bibfnamefont{K.}~\bibnamefont{Beloy}},
  \bibinfo{journal}{Phys. Rev. Lett.} \textbf{\bibinfo{volume}{112}},
  \bibinfo{pages}{062503} (\bibinfo{year}{2014}).

\bibitem[{\citenamefont{Seif and Mansour}(2015)}]{Seif-15}
\bibinfo{author}{\bibfnamefont{W.~M.} \bibnamefont{Seif}} \bibnamefont{and}
  \bibinfo{author}{\bibfnamefont{H.}~\bibnamefont{Mansour}},
  \bibinfo{journal}{Int. J. Mod. Phys. E} \textbf{\bibinfo{volume}{24}},
  \bibinfo{pages}{1550083} (\bibinfo{year}{2015}).

\bibitem[{\citenamefont{Bohr and Weisskopf}(1950)}]{Bohr-50}
\bibinfo{author}{\bibfnamefont{A.}~\bibnamefont{Bohr}} \bibnamefont{and}
  \bibinfo{author}{\bibfnamefont{V.~F.} \bibnamefont{Weisskopf}},
  \bibinfo{journal}{Phys. Rev.} \textbf{\bibinfo{volume}{77}},
  \bibinfo{pages}{94} (\bibinfo{year}{1950}).

\bibitem[{\citenamefont{Bellac}(1963)}]{LeBellac-63}
\bibinfo{author}{\bibfnamefont{M.~Le} \bibnamefont{Bellac}},
  \bibinfo{journal}{Nucl. Phys.} \textbf{\bibinfo{volume}{40}},
  \bibinfo{pages}{645} (\bibinfo{year}{1963}).

\bibitem[{\citenamefont{Bohr and Mottelson}(1998)}]{Bohr-98-II}
\bibinfo{author}{\bibfnamefont{A.}~\bibnamefont{Bohr}} \bibnamefont{and}
  \bibinfo{author}{\bibfnamefont{B.~R.} \bibnamefont{Mottelson}},
  \emph{\bibinfo{title}{Nuclear Structure. Vol. II: Nuclear Deformations.}}
  (\bibinfo{publisher}{World Scientific}, \bibinfo{address}{London},
  \bibinfo{year}{1998}).

\bibitem[{\citenamefont{Helmer and Reich}(1994)}]{Helmer-94}
\bibinfo{author}{\bibfnamefont{R.~G.} \bibnamefont{Helmer}} \bibnamefont{and}
  \bibinfo{author}{\bibfnamefont{C.~W.} \bibnamefont{Reich}},
  \bibinfo{journal}{Phys. Rev. C} \textbf{\bibinfo{volume}{49}},
  \bibinfo{pages}{1845} (\bibinfo{year}{1994}).

\bibitem[{\citenamefont{Abramowitz and Stegun}(1964)}]{Abramowitz-64}
\bibinfo{author}{\bibfnamefont{M.}~\bibnamefont{Abramowitz}} \bibnamefont{and}
  \bibinfo{author}{\bibfnamefont{I.~A.} \bibnamefont{Stegun}},
  \emph{\bibinfo{title}{Handbook of Mathematical Functions}}
  (\bibinfo{publisher}{National Bureau of Standards},
  \bibinfo{address}{Washington, D.C.}, \bibinfo{year}{1964}).

\bibitem[{\citenamefont{Finkbeiner et~al.}(1993)\citenamefont{Finkbeiner,
  Fricke, and Kuhl}}]{Finkbeiner-93}
\bibinfo{author}{\bibfnamefont{M.}~\bibnamefont{Finkbeiner}},
  \bibinfo{author}{\bibfnamefont{B.}~\bibnamefont{Fricke}}, \bibnamefont{and}
  \bibinfo{author}{\bibfnamefont{T.}~\bibnamefont{Kuhl}},
  \bibinfo{journal}{Phys. Lett. A} \textbf{\bibinfo{volume}{176}},
  \bibinfo{pages}{113} (\bibinfo{year}{1993}).

\bibitem[{\citenamefont{C.~E.~Bemis et~al.}(1988)\citenamefont{C.~E.~Bemis,
  McGowan, J.~L. C.~Ford et~al.}}]{Bemis-88-s}
\bibinfo{author}{\bibfnamefont{J.}~\bibnamefont{C.~E.~Bemis}},
  \bibinfo{author}{\bibfnamefont{F.~K.} \bibnamefont{McGowan}},
  \bibinfo{author}{\bibfnamefont{J.}~\bibnamefont{J.~L. C.~Ford}},
  \bibnamefont{et~al.}, \bibinfo{journal}{Phys. Scr.}
  \textbf{\bibinfo{volume}{38}}, \bibinfo{pages}{657} (\bibinfo{year}{1988}).

\bibitem[{\citenamefont{Dykhne and Tkalya}(1998)}]{Dykhne-98_ME}
\bibinfo{author}{\bibfnamefont{A.~M.} \bibnamefont{Dykhne}} \bibnamefont{and}
  \bibinfo{author}{\bibfnamefont{E.~V.} \bibnamefont{Tkalya}},
  \bibinfo{journal}{JETP Lett.} \textbf{\bibinfo{volume}{67}},
  \bibinfo{pages}{251} (\bibinfo{year}{1998}).

\bibitem[{\citenamefont{Kroger and Reich}(1976)}]{Kroger-76}
\bibinfo{author}{\bibfnamefont{L.}~\bibnamefont{Kroger}} \bibnamefont{and}
  \bibinfo{author}{\bibfnamefont{C.}~\bibnamefont{Reich}},
  \bibinfo{journal}{Nucl. Phys. A} \textbf{\bibinfo{volume}{259}},
  \bibinfo{pages}{29} (\bibinfo{year}{1976}).

\bibitem[{\citenamefont{Tkalya}(1992)}]{Tkalya-92}
\bibinfo{author}{\bibfnamefont{E.~V.} \bibnamefont{Tkalya}},
  \bibinfo{journal}{Nucl. Phys. A} \textbf{\bibinfo{volume}{539}},
  \bibinfo{pages}{209} (\bibinfo{year}{1992}).

\bibitem[{\citenamefont{Tkalya}(1994)}]{Tkalya-94}
\bibinfo{author}{\bibfnamefont{E.~V.} \bibnamefont{Tkalya}},
  \bibinfo{journal}{JETP} \textbf{\bibinfo{volume}{78}}, \bibinfo{pages}{239}
  (\bibinfo{year}{1994}).

\bibitem[{\citenamefont{Tkalya et~al.}(2015)\citenamefont{Tkalya, Schneider,
  Jeet, and Hudson}}]{Tkalya-15}
\bibinfo{author}{\bibfnamefont{E.~V.} \bibnamefont{Tkalya}},
  \bibinfo{author}{\bibfnamefont{C.}~\bibnamefont{Schneider}},
  \bibinfo{author}{\bibfnamefont{J.}~\bibnamefont{Jeet}}, \bibnamefont{and}
  \bibinfo{author}{\bibfnamefont{E.~R.} \bibnamefont{Hudson}},
  \bibinfo{journal}{Phys. Rev. C} \textbf{\bibinfo{volume}{92}},
  \bibinfo{pages}{054324} (\bibinfo{year}{2015}).

\bibitem[{\citenamefont{Church and Weneser}(1960)}]{Church-60}
\bibinfo{author}{\bibfnamefont{E.~L.} \bibnamefont{Church}} \bibnamefont{and}
  \bibinfo{author}{\bibfnamefont{J.}~\bibnamefont{Weneser}},
  \bibinfo{journal}{Ann. Rev. Nucl. Sci.} \textbf{\bibinfo{volume}{10}},
  \bibinfo{pages}{193} (\bibinfo{year}{1960}).

\bibitem[{\citenamefont{Davydov}(1965)}]{Davydov-65}
\bibinfo{author}{\bibfnamefont{A.~S.} \bibnamefont{Davydov}},
  \emph{\bibinfo{title}{Quantum Mechanics}} (\bibinfo{publisher}{Pergamon
  Press}, \bibinfo{address}{Oxford, England}, \bibinfo{year}{1965}).

\bibitem[{\citenamefont{Winston}(1963)}]{Winston-63}
\bibinfo{author}{\bibfnamefont{R.}~\bibnamefont{Winston}},
  \bibinfo{journal}{Phys. Rev.} \textbf{\bibinfo{volume}{129}},
  \bibinfo{pages}{2766} (\bibinfo{year}{1963}).

\bibitem[{\citenamefont{Church and Weneser}(1956)}]{Church-56}
\bibinfo{author}{\bibfnamefont{E.~L.} \bibnamefont{Church}} \bibnamefont{and}
  \bibinfo{author}{\bibfnamefont{J.}~\bibnamefont{Weneser}},
  \bibinfo{journal}{Phys. Rev.} \textbf{\bibinfo{volume}{104}},
  \bibinfo{pages}{1382} (\bibinfo{year}{1956}).

\end{thebibliography}
\end{document}